A Novel Algorithm for the Maximal Fit Problem in Boolean Networks

Guy Karlebach[1]

[1]MIT-Broad Foundry, Broad Institute of MIT and Harvard, Cambridge,

Massachusetts, USA.

Email: guykarle@broadinstitute.org

**Abstract**

Gene regulatory networks (GRNs) are increasingly used for explaining biological processes with complex transcriptional regulation. A GRN links the expression levels of a set of genes via regulatory controls that gene products exert on one another. Boolean networks are a common modeling choice since they balance between detail and ease of analysis. However, even for Boolean networks the problem of fitting a given network model to an expression dataset is NP-Complete. Previous methods have addressed this issue heuristically or by focusing on acyclic networks and specific classes of regulation functions. In this paper we introduce a novel algorithm for this problem that makes use of sampling in order to handle large datasets. Our algorithm can handle time series data for any network type and steady state data for acyclic networks. Using in-silico time series data we demonstrate good performance on large datasets with a significant level of noise.

**keywords**: Boolean network, Inference, Sampling

**Introduction**

Numerous biological phenomena arise through interactions between cellular components[1]. Gene regulatory networks (GRNs) are a paradigm that explains various processes such as embryonic development, circadian rhythms and disease progression as a product of interactions between genes that regulate each other's expression levels [2-4]. Various methodologies were suggested for modeling and analyzing these networks [5].

One of the simplest GRN models is the Boolean network[6]. Gene expression levels are marked as either expressed (Boolean 1) or not expressed (Boolean 0), and regulatory interactions such as those performed by transcription factors are described using Boolean functions. Although this formulation is simple, it can characterize a broad range of networks and dynamic behaviors [7, 8].

Researchers have successfully used Boolean networks for establishing various biological hypotheses. For example, Marr et al. [9] showed that the steady states of their Boolean network correspond to the differentiation states of lymphocytes. Similarly, Orlando et al. showed that a Boolean network model can predict cell cycle states, and explain the cyclic gene expression patterns that they observed in their dataset [10]. There exist examples from a diverse range of systems, including sporulation in *B.subtillis* [11], tryptophan biosynthesis in *E.coli* [12], floral organ determination in *A.thaliana* [13], and more. Usually the network's states are not derived directly from the data, but rather are determined using independent analysis or simulation and then compared to the data.

A different approach derives the Boolean states of the network directly from the data, such that each measurement is assigned an inferred network state. This allows for example comparison of the network behavior in cases and controls, or of the differences in trajectories of wild-type and mutant strains [14]. Karlebach and Shamir showed that the problem of finding the best dataset fit to a Boolean network is NP-Complete [15], and therefore a solution that can handle every instance of the problem efficiently is not likely to exist. Our goal in this work is not to prove the biological merit of this approach since this has been done elsewhere [14-16], but to systematically investigate a proposed method for alleviating intractability in large datasets.

Several methods that apply to specific network types or greedily search for a solution have been developed. Karlebach and Shamir proposed an inference algorithm that gradually updates the belief in each Boolean value, and can be used with small datasets and uncertain network topologies [15]. Sharan and Karp used linear programming to solve the problem for acyclic networks in steady state, and showed that it performs well, in particular for a specific class of regulation functions, and that it reliably predicts the regulation functions of signaling networks [16]. Some methods exist for deriving Boolean states from the dataset alone, which can then be compared to a network model [17, 18]. In addition, there are various algorithms for computing the Boolean regulation functions [19-21].

In this paper we take a different approach by making assumptions about the nature of the noise. If the noise is not correlated with a specific regulation function, then as datasets become large, inferring individual network states is more difficult than inferring the types of regulatory interactions, since regulatory interactions occur repeatedly in the data, whereas individual states may occur just once or a few times. Based on this argument, we present a novel algorithm that uses random sampling in order to infer individual states. The algorithm is suitable for time series data or for steady state data in acyclic networks, and we demonstrate its performance using the former. The main idea behind the algorithm is finding a trajectory that fits the largest number of Boolean data points in a sample. With enough data points for the fit, the sample is sufficient for overcoming the effect of incorrect measurements and detecting the correct trajectory. Since going over all possible initial states is infeasible for medium or larger networks, we devise a method that can perform the fit without need of doing so. Our experiments also establish a link between the level of error in the data and the running time required for finding an optimal solution.

The paper is organized as follows: the next section defines the problem, describes the inference algorithm and explains how it achieves good performance and accuracy. The Testing section demonstrates the performance and accuracy of the algorithm using a large simulated dataset. Finally, in the Conclusion section we summarize our findings and outline future work.

Inference Algorithm

A Boolean network is a dynamic model that contains N nodes, which we will refer to as *genes*, and N Boolean functions, which we will refer to as *regulation functions*. The inputs of a regulation function are the Boolean values that are assigned to a subset of the genes (the genes in this subset are the *regulators*) at a time i, and the output of a function assigns a value to a single gene, called the **target**, at time i+1. A *state* is an assignment of Boolean values to all the genes at a given time point. An initial state assigns a Boolean value to every gene. Then, the states at subsequent times can be derived by simultaneous application of the regulation functions. A set of time consecutive states is known as a *trajectory* of the network (**Figure 1**). The graph that represents the relations between genes and their sets of regulators is called the *network topology*.

The state inference problem requires finding the correct trajectory of a given Boolean network model given a trajectory that contains errors, where errors are Boolean values that changed. In such a noisy trajectory, when the Boolean value of a data point is different than the output of its regulation function, we say that it constitutes a *discrepancy*. A Boolean value assigned to a specific gene at a specific time in the input trajectory will be referred to as a *data point*. Here we will assume that the changes are i.i.d random variables, in other probability of data point to correspond to an incorrect Boolean value is the same for every data point. The number of Boolean values in a trajectory of length T is N·T. In terms of computational complexity, neither T nor N is assumed to be constant, and therefore there are $2^{NT}$ different (noisy) trajectories for every network with N nodes and T time points.

The input for the maximal fit problem is a set of noisy trajectories and a network topology, and the output is a minimal set of changes that is necessary for eliminating all discrepancies. This objective is intended to reconstruct the original trajectories before noise was added. If the noise level is very high, e.g. every data point is flipped with probability 0.5, the problem is still defined, but the reconstruction will be meaningless.

Following is an outline for the suggested inference algorithm:

1. Infer the Boolean functions by selecting those functions that agree with the maximal number of states in the input trajectories.
2. For each input trajectory,
    a. Find the initial state that fits the largest number of data points in a random sample from this trajectory (**Figure 2**).
    b. If there are several initial states that fit the largest number of Boolean values, select the one that generates a trajectory with minimal difference from the input trajectory.
    c. Generate a trajectory starting from the selected initial state and return it as a solution for the corresponding input trajectory.

Step 1 can be performed using Branch & Bound, as described in [15]. In short, for each row in the truth table of a Boolean function, each one of the two alternative outputs is assigned a score that is proportional to the number of its occurrence in the data. Then, for selecting the best combination of outputs that constitute the Boolean function, we branch at each Boolean output, bound when the greedy, highest scoring completion of the sub-solution scores below the best solution found, and reject illegal solutions. A combination of outputs is an illegal solution if the function that it defines does not depend on the value of one of its regulators.

Assuming that the number of initial states returned by step **2a** is constant, Steps **2b** and **2c** are computed trivially in O(N·T), and so next we focus on step **2a**.

Denote a data point as (*g*,*t*,*b*), where *g* is some gene, *t* is some time and *b* is some Boolean value. An initial state *S* fits the data point (g,t,b) if the trajectory generated from *S* assigns the Boolean value *b* to gene *g* at time *t* (**Figure 2**). For each data point, a recursive strategy is used for finding all the initial states that fit it:

For a gene *g*, time *t* and Boolean value *b*:

1. For each combination of regulator values that generates *b*:
    a) Assign this combination of values to the regulators of *g* at time *t*-1
    b) Solve the problem recursively for each regulator
    c) Intersect the results returned from the recursive calls
2. Return a union of the results from step 1

The recursion returns a set that can contain an exponential number of states. However, its representation can be represented succinctly by marking genes that can take either Boolean value with a special character '?'. For example, if we have 3 genes, the following annotation: (1,?,0) represents two Boolean states: (1,0,0) and (1,1,0). All union and intersection operations can make use of this symbolic representation, and compress their results accordingly. For example, a union between states (1,?,?) and (1,1,?) can be represented by (1,?,?), and an intersection between these states by (1,1,?). We implement these operations using tries.

The stopping condition of the recursion occurs at the initial state, where the solution is trivial – all the states that contain a given set of data point values. Since a memoization table can be built once and used for every possible trajectory, even if this takes time exponential in N, for large T it is still constant per trajectory. In cases where the memorization table needs to be built frequently, for example if there is uncertainty about the network topology, the sets of states can be computed inaccurately. In other words, a larger set of states can be kept at each entry, such that its representation is smaller. This can be implemented in various ways. For the network in this study, the best tradeoff we found between preprocessing, running time and accuracy occurred when randomly replacing sub-tries that have over 100 leaves and represent a large number of sub-states by the full sub-state representation (a chain of ? nodes).

Given a representation of all the states for each data point, we can select an initial state that fits the largest number of sampled data points using Branch & Bound. We bound whenever the current sub-state is common to less data points than the value of the optimal solution found so far. An initial bound can be obtained as follows: Let $\pi$ be the proportion of erroneous data points in the sampled data points. A good value for an initial bound would be $(1-\pi) \cdot$(# sampled data points), because that is exactly the proportion of correct data points. However, since $\pi$ is not known, we will initialize $\pi$ to p, the probability of error, and use the following loop:

1. Perform B&B using $(1-p) \cdot$(# sampled data points) as an initial bound
2. Increase p by 0.05

Until a solution is found.

When the sample size is large enough, the initial state that we will fit to will be the same one that fits the largest number of data points in the complete trajectory. If in practice there is more than one initial state that fits them, we select as the solution the one that is most similar to the input trajectory. For the network in this study we observed that as T increases, data points become less informative about the initial state, and so we adjusted the sampling probability to decrease with T.

Note that step **2** of the algorithm can be parallelized, although this functionality is currently not implemented in our code. The next section describes the tests that we performed in order to confirm efficiency and accuracy of the inference algorithm.

**Testing**

Since the state inference problem is NP-Complete [15], an algorithm's ability to cope with large datasets is crucial for its general applicability. Our algorithm optimizes the same objective function as in [14-16], and therefore its usefulness in analyzing biological datasets follows directly from the findings of these studies. In order to demonstrate that it is suitable for larger datasets and more complex network topologies, we construct the following Boolean network: The network has 25 genes, each of which has 2 regulators (**Figure 3**). The regulators are chosen such that by iterating backwards from a gene to its regulator and to that regulator's regulator and so on, we can reach any other gene. This choice ensures that the network cannot be simplified into independent subnetworks, and that every pair of genes has the potential to influence one another in every trajectory. The regulation functions are XOR, as this choice produces complex dynamic behaviors.

The test dataset contains 100,000 data points, divided into 40 trajectories of length 100 each. A trajectory of length 100 of a cyclic network with 25 nodes has the same number of data points as a steady state acyclic network with 2,500 nodes. The following probabilities are used for generating errors in the input dataset: $p_1$=0.05, $p_2$=0.1, $p_3$=0.15 and $p_4$=0.2. For example, when using $p_4$ we change on average every fifth Boolean value in the input. The effects of these noise levels on the first 15 states of the first trajectory in the dataset are illustrated in **Figure 4**.

Then, for each probability $p_i$ we measure the running time of the algorithm and the number of incorrectly inferred Boolean values. **Figure 5** summarizes the performance of the algorithm using these error levels. As can be seen in the figure, increasing the error level increases the running time and reduces the accuracy of prediction. This is to be expected since higher error levels mean that more flipped data points are sampled, and therefore finding an initial state that is shared between the maximal number of points is harder. For each sample size and noise level, we computed a new memorization table and included its construction time in the average running time per trajectory.

So far we assumed that the probability of error of data points is the same. In order to test a different pattern of noise, we now define a probability of error $p_t$ as follows:

$$p_t = \begin{cases} 0 & t \in \{1,3,5 \ldots\} \\ 0.3 & t \in \{2,4,6 \ldots\} \end{cases}$$

where t corresponds to time. At even times the probability $p_t$ that a Boolean value be incorrect is 0.3, and at odd it is 0. Therefore, the overall number of errors for $p_t$ is the same as $p_3$, but there is an association between time and error. **Figure6A** illustrates the effect of this noise scheme on the first 15 states of the first trajectory in the dataset. **Figure 6B** shows the number of mistakes the algorithm makes for different sample sizes, and **Figure 6C** the running time as a function of sample size. As can be seen in the figure, the algorithm makes more mistakes at small sample sizes than it makes for noise level $p_3$, which indicates that the association of noise with time makes data points with incorrect values share more initial states. Nevertheless, when the sample size is large enough the algorithm does not make any mistakes. The running times of the algorithm are shorter, most likely since this noise pattern creates a favorable search space for the branch and bound step of the algorithm.

For executing our program we used a Macbook air with a 2.2 GHz Intel Core i7 processor with 8 GB of memory. For optimal performance, we implemented the algorithm in C. The binary, source code and input files that were used in this study can be obtained for free by contacting the author.

**Conclusion**

Network models have been shown to agree well with observed patterns of gene expression. Currently, a gold-standard methodology for generating hypotheses about a network model given a dataset of gene expression does not exist. An important aspect in quantifying the usefulness of a given model is the analytical tools that are available when one adopts the model. Boolean networks are expressive, which means that they can describe a broad range of observations, and at the same time they are simple. Due to the latter property, inference algorithms can be developed and studied using existing theory [15, 16, 22-24].

In this paper we showed that very large datasets can be used for accurate inference despite the computational complexity of the problem. Our algorithm offers researchers a powerful tool for exploring network hypotheses and provides an incentive for generating large datasets. In addition, it provides insight into the network inference problem that can be used for development of new analysis methods.

There are several research directions that we plan to pursue in future work. First, The minimal amount of data that is needed in order to reconstruct a trajectory is an important quantity both for inference and for designing biological experiments. This includes the minimal trajectory length T, and the minimal number of trajectories in the input. It may be possible to derive the information dynamically when constructing the state memorization table. In addition, we have observed that for the XOR network data points with larger time T are less useful for inferring the initial state, and it is of interest to quantify this property. Another interesting question is what is the maximal level of noise that can be corrected. This level may depend on the dataset and network topology, and so a related question is whether there are trajectories or network topologies that are more robust to noise. Given a dataset, it is also desirable to find bounds on the number of changes needed to remove all discrepancies. A lower bound clearly exists, since a single change in a data point value cannot solve more discrepancies than one plus the number of targets of a gene. We believe that answers to these questions will be important for the understanding of complex systems.

**Figure 1**: A Boolean network model and one trajectory. The leftmost table represents the trajectory, with the initial state (1,0,0). The middle diagram is the network topology. It shows the regulators of each gene, where there is a directed edge from every regulator to its target. In this case, A regulates B, B regulates C, and C regulates A. Since the trajectory is noiseless, by comparing it with the network topology it is easy to infer that the regulation functions that determine the values of B and C are identity functions, whereas the function that determines A is negation. The tables on the rightmost column specify the model's regulation functions.

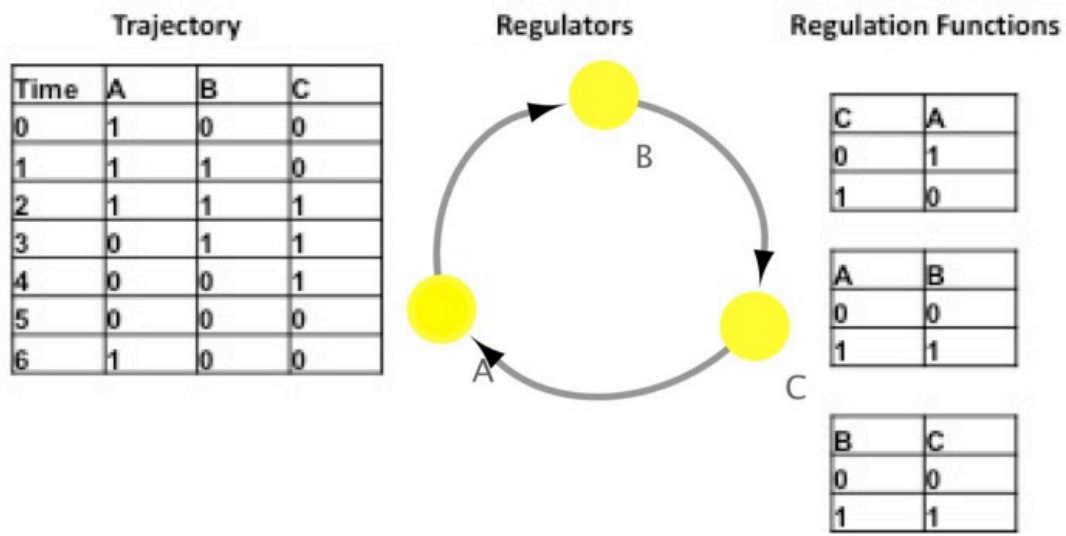

**Figure 2**: Intersecting sets of states in order to find an initial state that is common to as many sampled data points as possible (2 in this example). Given a network topology, a memorization table that encodes sets of initial states that lead to each sampled data point value is generated. Then, the corresponding sets of initial states (represented in this figure both as tries, in which '?' means "0 or 1", and as strings) are intersected. The memorization table can also store sets that only <u>contain</u> the source states of each data point value, as long as the intersection of the sampled data point states will retrieve an initial state that solves the maximal fit problem. Note that some entries of the memorization table contain two sets, one corresponding to a Boolean 0 and one to a Boolean 1. This can occur during the recursive construction of the table.

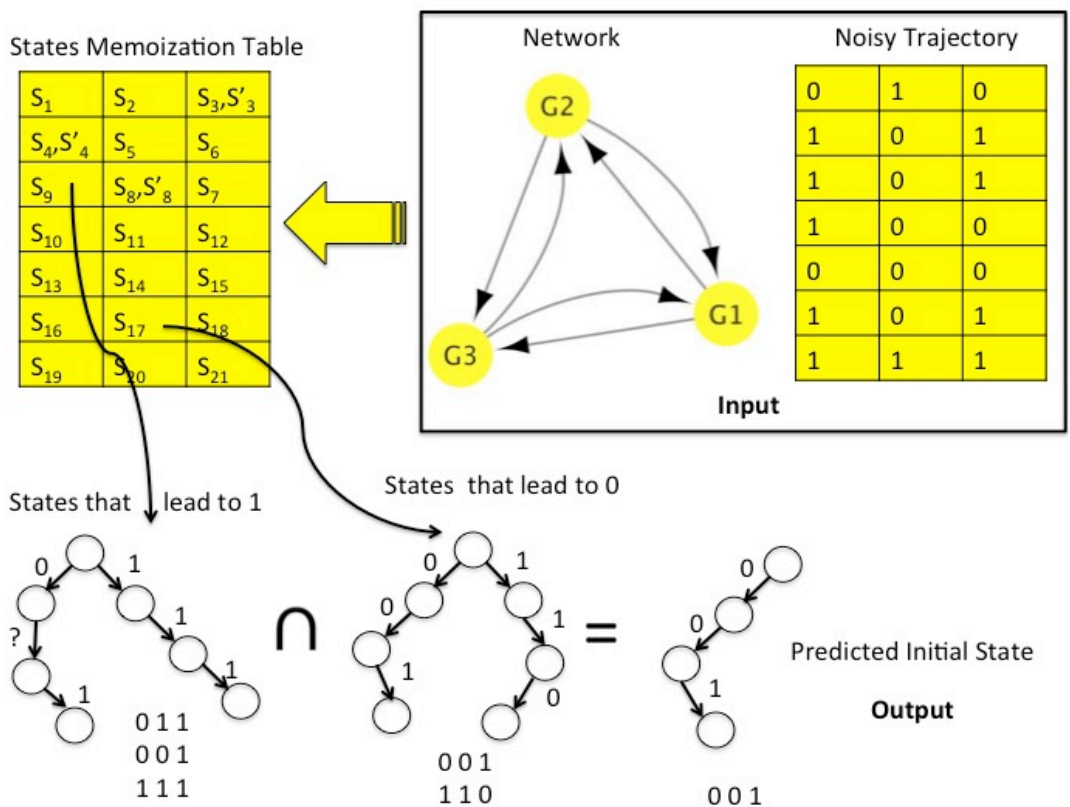

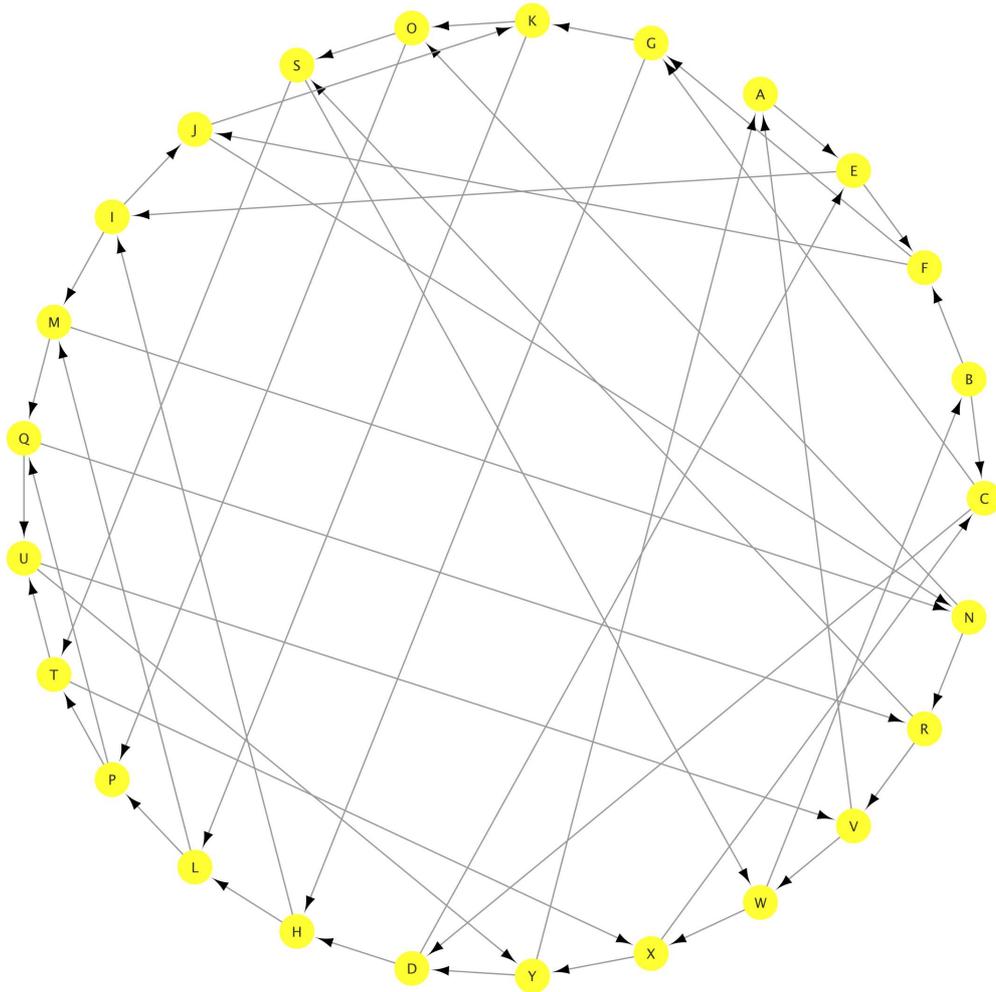

**Figure 3**: A diagram of the gene network that was used for testing the inference algorithm. The genes are drawn as circles, and there is a directed edge between every regulator and its target. The network has 25 genes, and each gene has two other genes as regulators, and is itself a regulator of another genes.

**Figure 4**: Illustration of the effect of noise on the first 15 states in the test network's first trajectory. The original states (blue) and the states with addition of noise (red) are subjected to multidimensional scaling, where the distance function is the number of different Boolean values between a pair of states. The states are numbered according to their order in the trajectory. When the level of noise is 0.05 (top left frame), corresponding noisy and noiseless states are relatively close, or even identical in the case of states 2, 7 and 10. As noise level increases the distances between corresponding states grow.

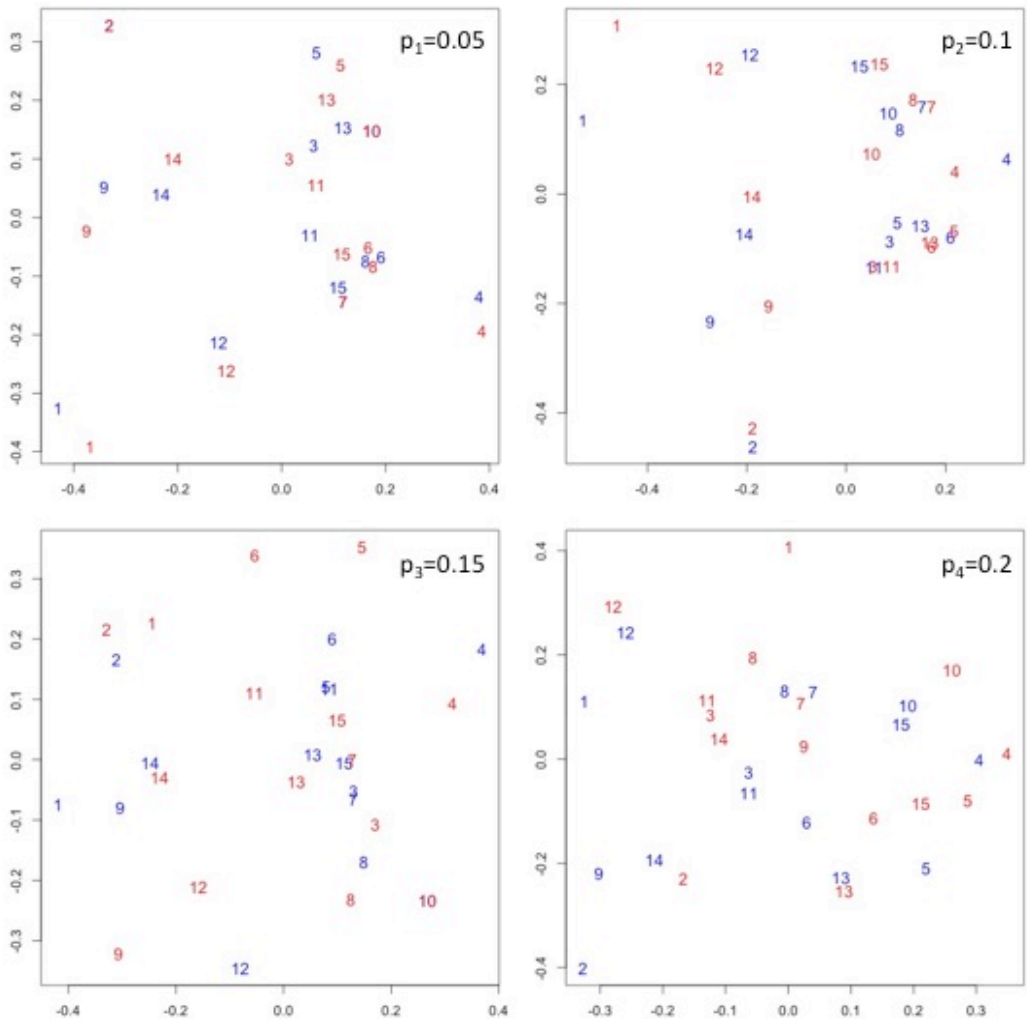

**Figure 5:** **A:** The percentage of incorrectly classified data points as a function of sample size, for the four different noise levels tested in the paper. For error levels $p_1=0.05$ (magenta) and $p_2=0.1$ (blue), all sample sizes result in mistake-free reconstruction. The higher the noise level, the larger the sample size needed for eliminating errors. **B:** Running times of the algorithm for different sample sizes and noise levels. The y-axis is in log(seconds) scale for display purposes. Running times consistently increase with noise level for all sample sizes.

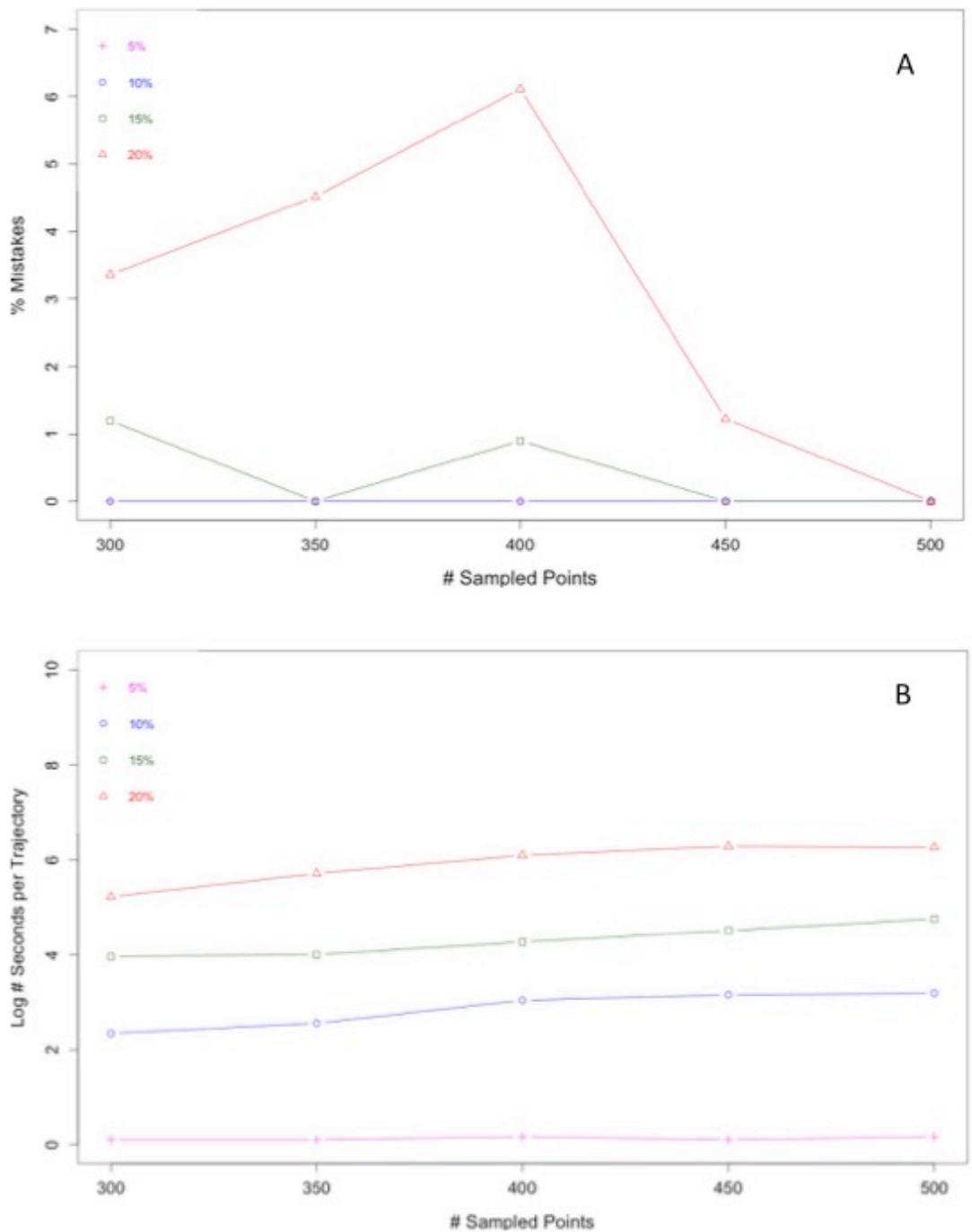

**Figure 6:** **A:** Multidimensional scaling of the first 15 states in the first trajectory in the dataset, without noise (blue) and time-correlated noise (also see text). The distance function is the number of different Boolean values between a pair of states. The states are numbered according to their order in the trajectory. The first state and every odd state are noiseless, and therefore the blue and red odd numbers are completely overlapping. The other time points contain an error with probability 0.4 **B:** The percentage of incorrectly classified data points as a function of sample size, for the time-correlated error level $p_t$. **C:** Running times of the algorithm for different sample sizes, measured in seconds.

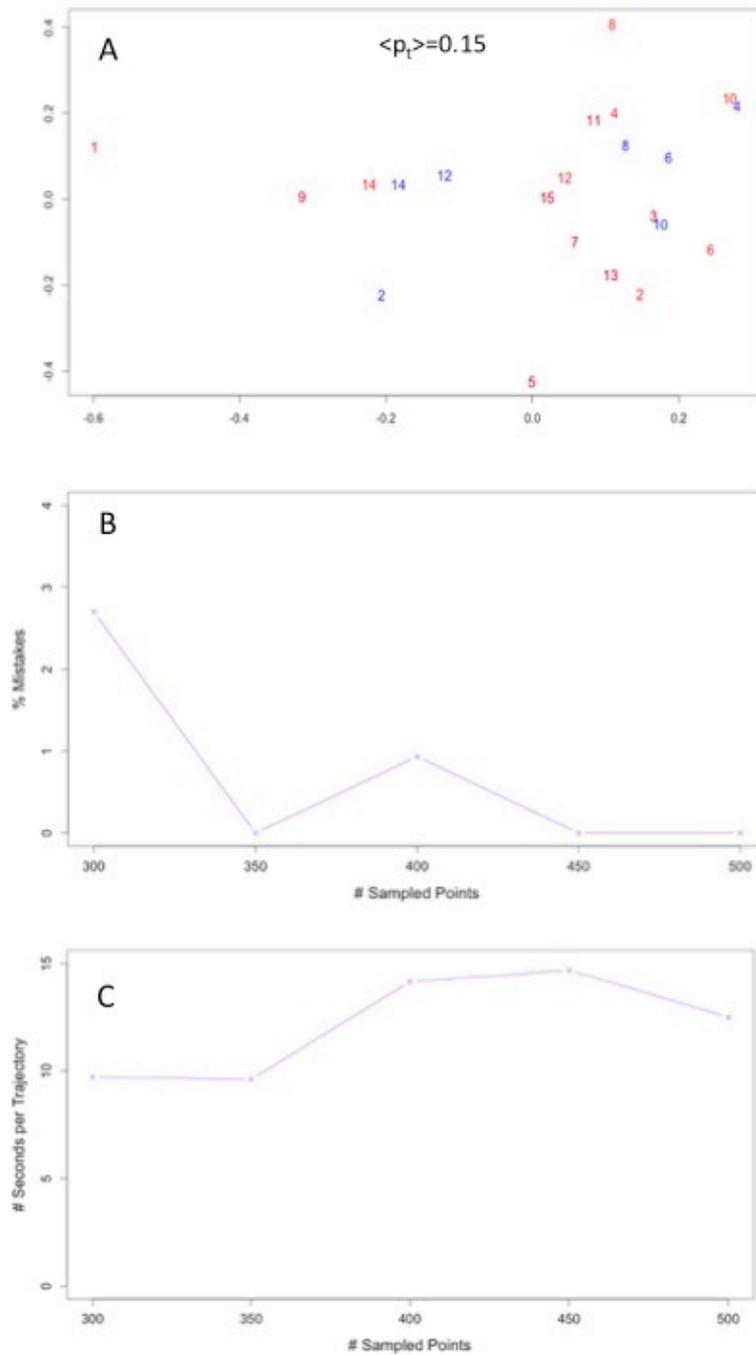

**Competing Interests**

The author declares that he has no competing interests